\documentclass[12pt]{article}
\usepackage{a4wide,epsfig,psfrag,amsmath,amssymb,cite,scalefnt}
\usepackage{color}

\usepackage{axodraw4j}

\newcommand{\qsla}{q\!\!\!/\,\,}
\newcommand{\Qsla}{Q\!\!\!\!/\,\,\,}

\newcommand{\gsim}{\;\rlap{\lower 3.5 pt \hbox{$\mathchar \sim$}} \raise 1pt
 \hbox {$>$}\;}
\newcommand{\lsim}{\;\rlap{\lower 3.5 pt \hbox{$\mathchar \sim$}} \raise 1pt
 \hbox {$<$}\;}

\allowdisplaybreaks

\begin{document}

\title{\vskip-3cm{\baselineskip14pt
    \begin{flushleft}
      \normalsize SFB/CPP-13-03\\
      \normalsize TTP13-02 
  \end{flushleft}}
  \vskip1.5cm
  Four-loop corrections with two closed fermion loops to
  fermion self energies and the lepton anomalous magnetic moment
}

\author{
  Roman Lee$^{a}$,
  Peter Marquard$^{b}$,
  Alexander V. Smirnov$^{c}$,
  \\
  Vladimir A. Smirnov$^{d,e}$,
  Matthias Steinhauser$^{b}$
  \\[1em]
  {\small\it (a) Budker Institute of Nuclear Physics and Novosibirsk State
    University}\\
  {\small\it 630090 Novosibirsk, Russia}
  \\
  {\small\it (b) Institut f{\"u}r Theoretische Teilchenphysik,}
  {\small\it Karlsruhe Institute of Technology (KIT)}\\
  {\small\it 76128 Karlsruhe, Germany}
  \\
  {\small\it (c) Scientific Research Computing Center,}
  {\small\it Moscow State University}\\
  {\small\it 119992 Moscow, Russia}
  \\
  {\small\it (d) Skobeltsyn Institute of Nuclear Physics, Moscow State University}\\
  {\small\it 119992 Moscow, Russia}
  \\
  {\small\it (e) Institut f\"{u}r Mathematik, Humboldt-Universit\"{a}t zu Berlin}\\
  {\small\it 12489 Berlin, Germany}
}

\date{}

\maketitle

\thispagestyle{empty}

\begin{abstract}
  We compute the eighth-order 
  fermionic corrections involving two and three closed massless
  fermion loops to the anomalous magnetic moment of the muon.
  The required four-loop on-shell integrals are classified and explicit
  analytical results for the master integrals are presented.
  As further applications we compute the corresponding 
  four-loop QCD corrections to the mass
  and wave function renormalization constants for a massive quark 
  in the on-shell scheme.
  \medskip

  \noindent
  PACS numbers: 12.20.Ds 12.38.Bx 14.65.-q 
\end{abstract}

\thispagestyle{empty}

%- }}}

\newpage

%- {{{ Introduction:

\section{Introduction}

In the last about ten years several groups have been active in
computing four-loop corrections to various physical quantities.  Among
them are the order $\alpha_s^4$ corrections to the $R$
ratio and the Higgs decay into bottom
quarks~\cite{Baikov:2005rw,Baikov:2008jh,Baikov:2012er}, four-loop corrections
to moments of the photon polarization
function\cite{Chetyrkin:2006xg,Boughezal:2006px,Sturm:2008eb,Maier:2008he,Maier:2009fz}
which lead to precise results for the charm and bottom quark masses
(see, e.g., Ref.~\cite{Chetyrkin:2009fv}), and the free energy density
of QCD at high temperatures~\cite{Kajantie:2002wa}.  The integrals
involved in such calculations are either four-loop massless two-point
functions or four-loop vacuum integrals with one non-vanishing mass
scale.  In this paper we take the first steps towards the systematic
study of a further class of four-loop single-scale integrals, the
so-called on-shell integrals where in the loop massless and massive
propagators may be present and the only external momentum is on the
mass shell.

\begin{figure}[t]
\SetScale{0.7}

\begin{center}
\fcolorbox{white}{white}{
  \begin{picture}(707,276) (48,-100)
%  \begin{picture}(567,392) (80,-123)
    \SetWidth{2.0}
    \SetColor{Black}
    \Arc[arrow,arrowpos=0.63,arrowlength=12.5,arrowwidth=5,arrowinset=0.2](351,114)(41.012,91,451)
    \Arc[arrow,arrowpos=0.63,arrowlength=12.5,arrowwidth=5,arrowinset=0.2](349,114)(16,90,450)
    \Line[arrow,arrowpos=0.5,arrowlength=12.5,arrowwidth=5,arrowinset=0.2](560,8)(592,-40)
    \Line[arrow,arrowpos=0.5,arrowlength=12.5,arrowwidth=5,arrowinset=0.2](592,-40)(528,-40)
    \Line[arrow,arrowpos=0.5,arrowlength=12.5,arrowwidth=5,arrowinset=0.2](496,-104)(624,-104)
    \Line[arrow,arrowpos=0.5,arrowlength=12.5,arrowwidth=5,arrowinset=0.2](624,-104)(640,-120)
    \Photon(624,-104)(592,-40){7.5}{4}
    \Line[arrow,arrowpos=0.5,arrowlength=12.5,arrowwidth=5,arrowinset=0.2](528,-40)(560,8)
    \Line[arrow,arrowpos=0.5,arrowlength=12.5,arrowwidth=5,arrowinset=0.2](480,-120)(496,-104)
    \Photon(560,8)(560,40){7.5}{2}
    \Arc[arrow,arrowpos=0.5,arrowlength=12.5,arrowwidth=5,arrowinset=0.2](512,-72)(16,90,450)
    \Photon(496,-104)(512,-88){7.5}{1}
    \Photon(512,-56)(528,-40){7.5}{1}
    \Photon(560,-104)(560,-40){7.5}{4}
    \Line[arrow,arrowpos=0.5,arrowlength=12.5,arrowwidth=5,arrowinset=0.2](97,-104)(193,8)
    \Line[arrow,arrowpos=0.5,arrowlength=12.5,arrowwidth=5,arrowinset=0.2](193,8)(289,-104)
    \Photon(193,8)(193,40){7.5}{2}
    \Photon(108,-95)(128,-93){7.5}{1}
    \Arc[arrow,arrowpos=0.5,arrowlength=12.5,arrowwidth=5,arrowinset=0.2](143,-95)(16,90,450)
    \Arc[arrow,arrowpos=0.5,arrowlength=12.5,arrowwidth=5,arrowinset=0.2](191,-95)(16,90,450)
    \Arc[arrow,arrowpos=0.5,arrowlength=12.5,arrowwidth=5,arrowinset=0.2](239,-95)(16,90,450)
    \Photon(175,-95)(159,-95){7.5}{1}
    \Photon(223,-95)(207,-95){7.5}{1}
    \Photon(254,-96)(276,-91){7.5}{1}
    \Line[arrow,arrowpos=0.5,arrowlength=12.5,arrowwidth=5,arrowinset=0.2](472,101)(552,213)
    \Line[arrow,arrowpos=0.5,arrowlength=12.5,arrowwidth=5,arrowinset=0.2](552,213)(632,101)
    \Photon(552,213)(552,245){7.5}{2}
    \PhotonArc[clock](580.855,170.855)(30.272,123.834,21.603){7.5}{3.5}
    \Photon(494,134)(534,135){7.5}{2}
    \Arc[arrow,arrowpos=0.5,arrowlength=12.5,arrowwidth=5,arrowinset=0.2](552,136)(16,90,450)
    \Arc[arrow,arrowpos=0.715,arrowlength=12.5,arrowwidth=5,arrowinset=0.2](622,171)(16.279,313,673)
    \Photon(608,136)(569,137){7.5}{2}
    \Line[arrow,arrowpos=0.5,arrowlength=12.5,arrowwidth=5,arrowinset=0.2](304,-104)(384,8)
    \Line[arrow,arrowpos=0.5,arrowlength=12.5,arrowwidth=5,arrowinset=0.2](384,8)(464,-104)
    \Photon(384,8)(384,40){7.5}{2}
    \PhotonArc[clock](405.258,-68.027)(50.078,92.584,-21.032){7.5}{5.5}
    \Photon(326,-71)(346,-69){7.5}{1}
    \Arc[arrow,arrowpos=0.5,arrowlength=12.5,arrowwidth=5,arrowinset=0.2](361,-71)(16,90,450)
    \Arc[arrow,arrowpos=0.5,arrowlength=12.5,arrowwidth=5,arrowinset=0.2](409,-71)(16,90,450)
    \Photon(393,-71)(377,-71){7.5}{1}
    \Photon(441,-71)(425,-71){7.5}{1}
    \Line[arrow,arrowpos=0.5,arrowlength=12.5,arrowwidth=5,arrowinset=0.2](257,104)(353,216)
    \Line[arrow,arrowpos=0.5,arrowlength=12.5,arrowwidth=5,arrowinset=0.2](353,216)(449,104)
    \Photon(353,216)(353,248){7.5}{2}
    \Photon(268,113)(308,117){7.5}{2}
    \Photon(348,96)(353,73){7.5}{1}
    \Photon(348,153)(351,127){7.5}{2}
    \Photon(390,113)(436,117){7.5}{2}
    \Line[arrow,arrowpos=0.5,arrowlength=12.5,arrowwidth=5,arrowinset=0.2](96,104)(160,216)
    \Line[arrow,arrowpos=0.5,arrowlength=12.5,arrowwidth=5,arrowinset=0.2](160,216)(224,104)
    \Photon(115,134)(204,134){7.5}{5}
    \Photon(160,216)(160,248){7.5}{2}
    \SetWidth{1.5}
    \Line[arrow,arrowpos=0.5,arrowlength=10,arrowwidth=4,arrowinset=0.2](176,264)(176,232)
    \Line[arrow,arrowpos=0.5,arrowlength=10,arrowwidth=4,arrowinset=0.2](224,136)(237,115)
    \Line[arrow,arrowpos=0.5,arrowlength=10,arrowwidth=4,arrowinset=0.2](82,113)(96,136)
    \Text(130,160)[lb]{\Large{\Black{$q$}}}
    \Text(45,92)[lb]{\Large{\Black{$p_1$}}}
    \Text(170,92)[lb]{\Large{\Black{$p_2$}}}
%    \Text(192,248)[lb]{\Large{\Black{$q$}}}
%    \Text(112,104)[lb]{\Large{\Black{$p_1$}}}
%    \Text(176,104)[lb]{\Large{\Black{$p_2$}}}
    \SetWidth{2.0}
    \PhotonArc[clock](612.248,137.762)(24.763,47.432,-66.809){7.5}{2.5}
  \end{picture}

}
\end{center}

  \caption[]{\label{fig::diags_gm2}
    Sample Feyman diagrams for the photon-muon vertex
    contributing to $a_\mu$. 
    Wavy and straight lines represent photons and fermions, respectively.
    In this paper we consider the contribution where at least two
    of the closed loops correspond to massless fermions.
    The last diagram in the second line is a representative
    of the so-called ``light-by-light'' contribution.
  }
\end{figure}
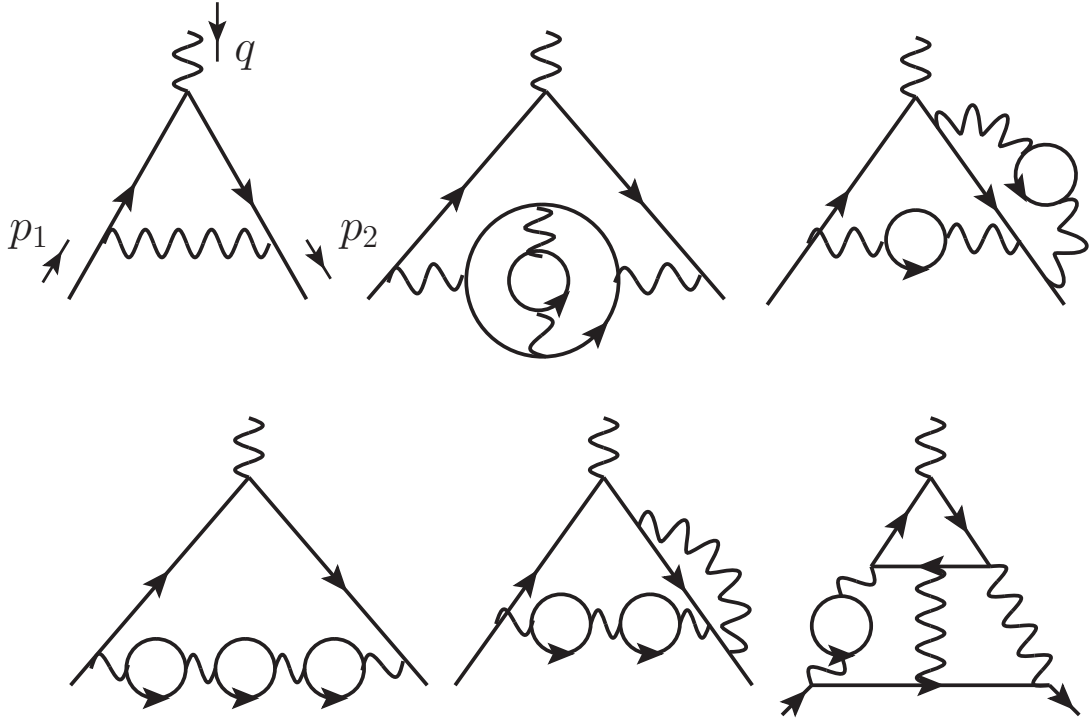

On-shell integrals enter a variety of physical quantities, where the
anomalous magnetic moments and on-shell counterterms are prominent
examples.  The first systematic study of two-loop on-shell integrals
needed for the evaluation of the on-shell mass and wave function
renormalization constants ($Z_m^{\rm OS}$ and $Z_2^{\rm OS}$) for a
heavy quark in QCD has been performed in
Refs.~\cite{Gray:1990yh,Broadhurst:1991fy}. Already a few years later,
in 1996 the analytical three-loop corrections to the lepton anomalous
magnetic moment $a_l$ became
available~\cite{Laporta:1996mq}. This result has been checked in
Refs.~\cite{Melnikov:2000qh,Marquard:2007uj}.  In
Refs.~\cite{Melnikov:2000qh,Melnikov:2000zc} the three-loop on-shell
integrals have been applied to QCD, namely the evaluation of $Z_m^{\rm
  OS}$ and $Z_2^{\rm OS}$.  The calculation of
Ref.~\cite{Melnikov:2000qh} has confirmed the numerical result
of~\cite{Chetyrkin:1999ys,Chetyrkin:1999qi} which has been available
before.  Both $Z_m^{\rm OS}$ and $Z_2^{\rm OS}$ have also been
computed in Ref.~\cite{Marquard:2007uj}.  Further application of
three-loop on-shell integrals are discussed in
Refs.~\cite{Melnikov:1999xp,Grozin:2007fh}. There is no systematic
study of four-loop on-shell integrals available in the
literature. 
Nevertheless, some four-loop results to the anomalous
magnetic moment of the muon, $a_\mu$, 
have been computed analytically, in particular contributions from closed 
electron loops. E.g., the contribution where the photon propagator
of the one-loop diagram (see Fig.~\ref{fig::diags_gm2}) 
is dressed by higher order corrections has been considered in several
papers~\cite{Lautrup:1974ic,Kinoshita:1990ur,Kawai:1991wd,Faustov:1990zs,Broadhurst:1992za,Baikov:1995ui,Baikov:2012rr}. Four-loop corrections where one of the
two photon propagators of the two-loop diagram is dressed by higher orders has
been considered in Ref.~\cite{Laporta:1993ds,Aguilar:2008qj}. Contributions
where both 
photon propagators get one-loop electron insertions are still missing.
This gap will be closed in the present work.
Let us mention that all four- and even
five-loop results for $a_l$ are available in the literature
in numerical
form~\cite{Kinoshita:2004wi,Aoyama:2007mn,Aoyama:2012wj,Aoyama:2012wk,Baikov:2012rr} (see also
the review articles~\cite{Jegerlehner:2008zza,Jegerlehner:2009ry}).

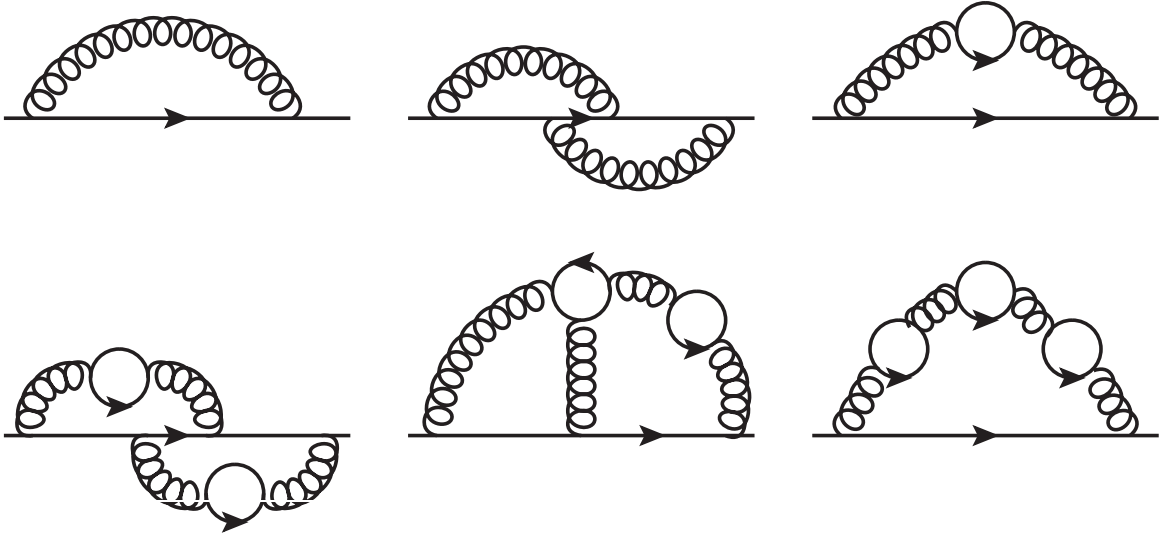
\begin{figure}[t]
\SetScale{0.68}

\begin{center}
\fcolorbox{white}{white}{
  \begin{picture}(644,196) (36,-60)
    \SetWidth{2.0}
    \SetColor{Black}
    \Line[arrow,arrowpos=0.5,arrowlength=12.5,arrowwidth=5,arrowinset=0.2](48,120)(240,120)
    \GluonArc(136,90.667)(77.746,22.166,157.834){7.5}{15}
    \Line[arrow,arrowpos=0.5,arrowlength=12.5,arrowwidth=5,arrowinset=0.2](272,120)(464,120)
    \GluonArc(336,100)(52,22.62,157.38){7.5}{10}
    \GluonArc[clock](400,140)(52,-22.62,-157.38){7.5}{10}
    \Line[arrow,arrowpos=0.5,arrowlength=12.5,arrowwidth=5,arrowinset=0.2](496,120)(688,120)
    \Line[arrow,arrowpos=0.5,arrowlength=12.5,arrowwidth=5,arrowinset=0.2](496,-56)(688,-56)
    \Arc[arrow,arrowpos=0.5,arrowlength=12.5,arrowwidth=5,arrowinset=0.2](544,-8)(16,90,450)
    \Arc[arrow,arrowpos=0.5,arrowlength=12.5,arrowwidth=5,arrowinset=0.2](592,24)(16,90,450)
    \Arc[arrow,arrowpos=0.5,arrowlength=12.5,arrowwidth=5,arrowinset=0.2](640,-8)(16,90,450)
    \Arc[arrow,arrowpos=0.5,arrowlength=12.5,arrowwidth=5,arrowinset=0.2](592,168)(16,90,450)
    \GluonArc(616,48)(126.491,108.435,145.305){7.5}{7}
    \GluonArc(568,48)(126.491,34.695,71.565){7.5}{7}
    \Line[arrow,arrowpos=0.5,arrowlength=12.5,arrowwidth=5,arrowinset=0.2](48,-56)(240,-56)
    \GluonArc(88,-48)(25.298,71.565,198.435){7.5}{5}
    \Arc[arrow,arrowpos=0.5,arrowlength=12.5,arrowwidth=5,arrowinset=0.2](112,-24)(16,90,450)
    \GluonArc(136,-48)(25.298,-18.435,108.435){7.5}{5}
    \Arc[arrow,arrowpos=0.5,arrowlength=12.5,arrowwidth=5,arrowinset=0.2](176,-88)(16,90,450)
    \GluonArc(152,-64)(25.298,161.565,288.435){7.5}{5}
    \GluonArc(200,-64)(25.298,-108.435,18.435){7.5}{5}
    \Line[arrow,arrowpos=0.705,arrowlength=12.5,arrowwidth=5,arrowinset=0.2](272,-56)(464,-56)
    \Arc[arrow,arrowpos=1,arrowlength=12.5,arrowwidth=5,arrowinset=0.2](368,24)(16,90,450)
    \GluonArc(396,1.438)(25.555,34.739,118.007){7.5}{3}
    \GluonArc(371.429,-57.143)(83.436,103.465,179.215){7.5}{9}
    \Arc[arrow,arrowpos=0.5,arrowlength=12.5,arrowwidth=5,arrowinset=0.2](432,8)(16,90,450)
    \GluonArc(414.939,-35.65)(38.822,-31.614,49.794){7.5}{5}
    \Gluon(368,-56)(368,8){7.5}{6}
    \Gluon(512,-56)(533,-19){7.5}{3}
    \Gluon(550,4)(576,24){7.5}{3}
    \Gluon(608,24)(626,-1){7.5}{2}
    \Gluon(652,-20)(672,-56){7.5}{3}
  \end{picture}
}
\end{center}

  \caption[]{\label{fig::diags_Zm}
    Sample Feynman diagrams for the QCD corrections to the fermion propagator
    contributing to $Z_m^{\rm OS}$ and $Z_2^{\rm OS}$.
    Curly and straight lines represent gluons and fermions, respectively.
    In this paper we consider the contribution where at least two
    of the closed loops correspond to massless fermions.
  }
\end{figure}

In this paper we take the first step towards the analytical calculation of
four-loop on-shell integrals by considering the subclass with
two or three closed massless fermion loops, which 
are marked by a factor $n_l$. Thus we are concerned with four-loop
terms proportional to $n_l^3$ and $n_l^2$ which we 
consider for
three physical quantities: the anomalous magnetic moment of the muon, $a_\mu$,
the on-shell mass renormalization constant, $Z_m^{\rm OS}$, and the on-shell
wave function renormalization constant, $Z_2^{\rm OS}$, for a massive
quark. For the latter QCD corrections to the quark two-point functions are
computed whereas for the former muon-photon vertex diagrams have to be
considered. Some sample Feynman diagrams are given in
Figs.~\ref{fig::diags_gm2} and~\ref{fig::diags_Zm}.
The precise definition of these quantities is provided in
Sections~\ref{sec::ZmZ2} and~\ref{sec::gm2}.

The outline of the paper is as follows: in the next section we provide details
of the four-loop on-shell integrals needed for our calculation. In particular,
we identify all master integrals and provide analytical results in Appendix~A.
The renormalization constants $Z_m^{\rm OS}$ and $Z_2^{\rm OS}$ are discussed
in Section~\ref{sec::ZmZ2} and Section~\ref{sec::gm2} is devoted to the
anomalous 
magnetic moment of the muon. We discuss the relation between the
$\overline{\rm MS}$ and on-shell fine structure constant and provide
analytical results for $a_\mu$.  Finally, we conclude in
Section~\ref{sec::concl}.
Appendix~B contains the analytic results for the relation between the 
fine structure constant defined in the $\overline{\rm MS}$ and on-shell
scheme. 

%- }}}
%- {{{ Four-loop on-shell integrals:

\section{\label{sec::os-ints}Four-loop on-shell integrals}

In this Section we present the setup used for the calculation and
discuss the families of
four-loop on-shell integrals needed for the $n_l^2$ and $n_l^3$ corrections
for $Z_2^{\rm OS}$, $Z_m^{\rm OS}$ and $a_\mu$. Since all three cases
reduce to the calculation of corrections to the fermion propagator 
we consider in this Section the corresponding two-point function.

After the generation of the diagrams with {\tt QGRAF}~\cite{Nogueira:1991ex}
we use {\tt q2e}~\cite{Harlander:1997zb,Seidensticker:1999bb} to translate the
output into a {\tt FORM}~\cite{Vermaseren:2000nd} readable form. In a next
step {\tt exp}~\cite{Harlander:1997zb,Seidensticker:1999bb} is applied
to map the momenta to one of five families. 
During the evaluation of the {\tt FORM} code we apply
projectors and take traces to end up with integrals which only contain
scalar products in the numerator and quadratic denominators.

In the next step we have to reduce all occurring integrals to a minimal set of
master integrals. This is done using two different programs in order to have a
cross check for the calculation. On the one hand we use {\tt
  crusher}~\cite{crusher} and on the other hand the {\tt C++} version of {\tt
  FIRE}.\footnote{The {\tt Mathematica} version of {\tt FIRE} is publicly
  available~\cite{Smirnov:2008iw}.} 
Both programs implements Laporta's algorithm~\cite{Laporta:2001dd} for
the solution of integration-by-parts identities~\cite{Chetyrkin:1981qh}.
We find complete agreement for the
expressions where the physical quantities are expressed in terms of master
integrals.

Let us mention that we have performed our calculations for general gauge
parameter which drops out once the four-loop results for $Z_2^{\rm OS}$,
$Z_m^{\rm OS}$ and $a_\mu$ are expressed in terms of master
integrals.\footnote{Note that $Z_m^{\rm OS}$ and $a_\mu$ have to be
  independent of the QCD gauge parameter $\xi$ whereas we expect that the
  $n_l^1$ and $n_l$-independent terms of $Z_2^{\rm OS}$ do depend on $\xi$.}

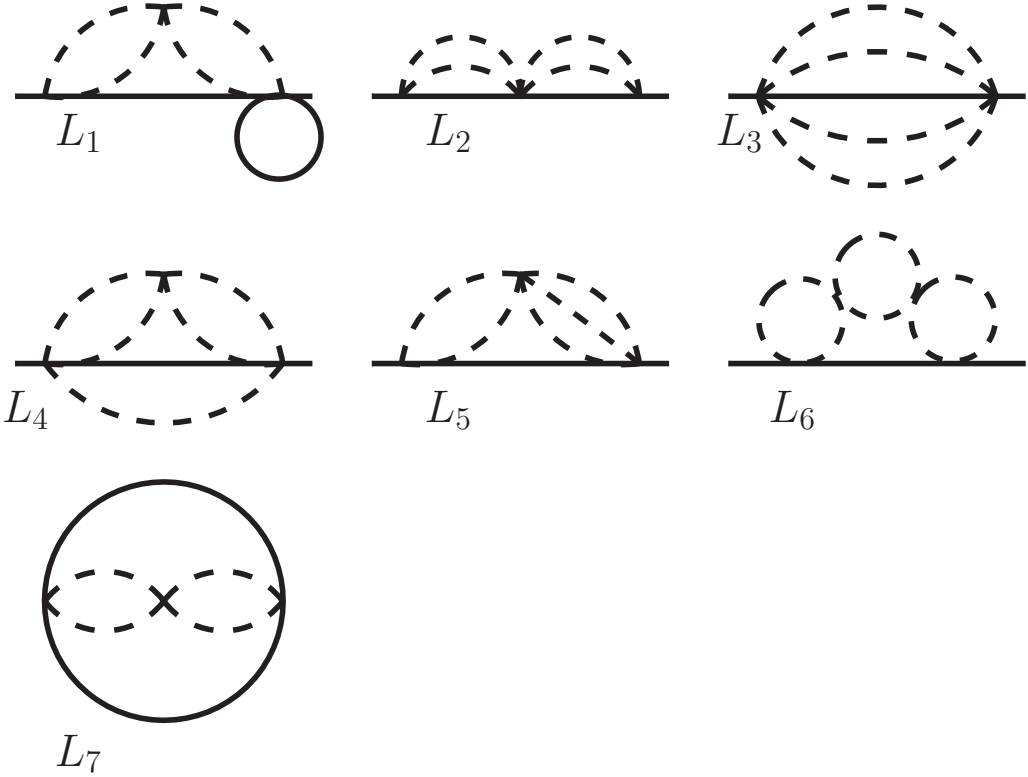
\begin{figure}[t]
\SetScale{0.7}

\begin{center}
\fcolorbox{white}{white}{
%%%  \begin{picture}(548,398) (62,-125)
  \begin{picture}(548,250) (42,-80)
    \SetWidth{3.0}
    \SetColor{Black}
    \Line(256,222)(416,222)
    \Arc[dash,dashsize=10,clock](368,198)(40,143.13,36.87)
    \Arc[dash,dashsize=10,clock](368,222)(32,-180,-360)
    \Arc[dash,dashsize=10,clock](304,222)(32,-180,-360)
    \Arc[dash,dashsize=10,clock](304,198)(40,143.13,36.87)
    \Line(448,222)(608,222)
    \Arc[dash,dashsize=10,clock](528,203.333)(66.667,163.74,16.26)
    \Arc[dash,dashsize=10,clock](528,148.667)(97.333,131.112,48.888)
    \Arc[dash,dashsize=10](528,240.667)(66.667,-163.74,-16.26)
    \Arc[dash,dashsize=10](528,295.312)(97.318,-131.12,-48.88)
    \Line(64,222)(224,222)
    \Arc[dash,dashsize=10,clock](152,214)(56.569,98.13,8.13)
    \Arc[dash,dashsize=10](200,278)(56.569,-171.87,-81.87)
    \Arc[dash,dashsize=10,clock](136,214)(56.569,171.87,81.87)
    \Arc[dash,dashsize=10](88,278)(56.569,-98.13,-8.13)
    \Arc(206,200)(22.627,135,495)
    \Line(448,78)(608,78)
    \Arc[dash,dashsize=10](487,101)(22.627,135,495)
    \Arc[dash,dashsize=10](528,125)(22.627,135,495)
    \Arc[dash,dashsize=10](569,102)(22.627,135,495)
    \Line(64,78)(224,78)
    \Arc[dash,dashsize=10,clock](152,70)(56.569,98.13,8.13)
    \Arc[dash,dashsize=10](200,134)(56.569,-171.87,-81.87)
    \Arc[dash,dashsize=10,clock](136,70)(56.569,171.87,81.87)
    \Arc[dash,dashsize=10](88,134)(56.569,-98.13,-8.13)
    \Arc[dash,dashsize=10](144,126)(80,-143.13,-36.87)
    \Line(256,78)(416,78)
    \Arc[dash,dashsize=10,clock](344,70)(56.569,98.13,8.13)
    \Arc[dash,dashsize=10](392,134)(56.569,-171.87,-81.87)
    \Arc[dash,dashsize=10,clock](328,70)(56.569,171.87,81.87)
    \Arc[dash,dashsize=10](280,134)(56.569,-98.13,-8.13)
    \Arc[dash,dashsize=10,clock](-415,-942)(1305.613,54.886,51.374)
    \Arc(144,-50)(64.203,143,503)
    \Arc[dash,dashsize=10,clock](112,-74)(40,143.13,36.87)
    \Arc[dash,dashsize=10](112,-26)(40,-143.13,-36.87)
    \Arc[dash,dashsize=10,clock](176,-74)(40,143.13,36.87)
    \Arc[dash,dashsize=10](176,-26)(40,-143.13,-36.87)
    \Text(60,135)[lb]{\Large{\Black{$L_1$}}}
    \Text(200,135)[lb]{\Large{\Black{$L_2$}}}
    \Text(310,135)[lb]{\Large{\Black{$L_3$}}}
    \Text(40,30)[lb]{\Large{\Black{$L_4$}}}
    \Text(200,30)[lb]{\Large{\Black{$L_5$}}}
    \Text(330,30)[lb]{\Large{\Black{$L_6$}}}
    \Text(60,-100)[lb]{\Large{\Black{$L_7$}}}
  \end{picture}
}
\end{center}

  \caption[]{\label{fig::MI1}
    Master integrals for the $n_l^2$ and $n_l^3$ contribution
    which are easily obtained by applying one- and two-loop
    formulae, see e.g., Ref.~\cite{Smirnov:2013}.
    Solid lines carry the mass $M$ and dashed lines are massless.
    For $L_1$ to $L_6$ we have $q^2=M^2$ where $q$ is the external
    momentum; $L_7$ is a vacuum integral.
  }
\end{figure}

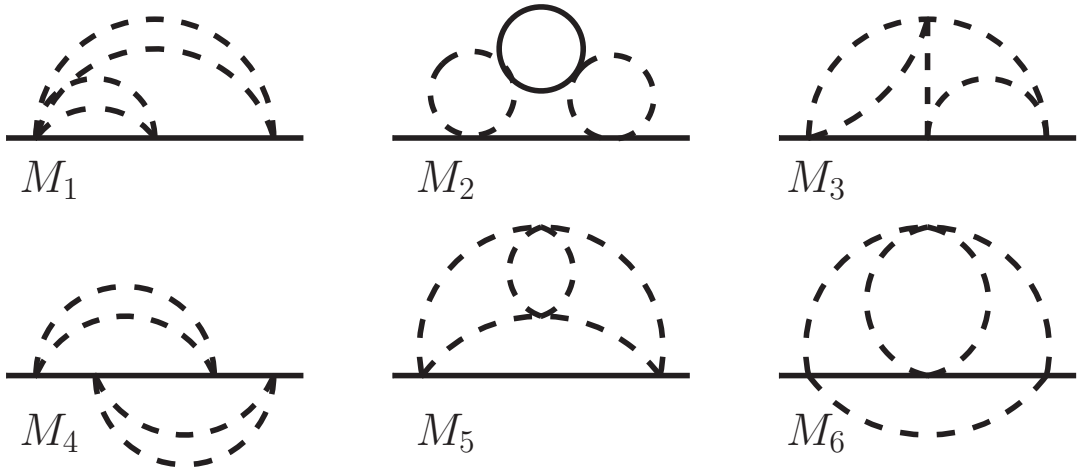
\begin{figure}[t]
\SetScale{0.7}
\begin{center}
\fcolorbox{white}{white}{
  \begin{picture}(580,180) (42,-73)
    \SetWidth{3.0}
    \SetColor{Black}
    \Line(480,84)(640,84)
    \Arc[dash,dashsize=10](472,172)(91.214,-74.745,-15.255)
    \Arc[dash,dashsize=10,clock](560,84)(64,-180,-360)
    \Arc[dash,dashsize=10,clock](592,84)(32,-180,-360)
    \Line[dash,dashsize=10](560,148)(560,84)
    \Line(272,84)(432,84)
    \Arc(352,132)(22.627,135,495)
    \Arc[dash,dashsize=10](390,106)(22.627,135,495)
    \Arc[dash,dashsize=10](315,108)(22.627,135,495)
    \Line(64,84)(224,84)
    \Arc[dash,dashsize=10,clock](144,84)(64,-180,-360)
    \Arc[dash,dashsize=10,clock](144,65.333)(66.667,163.74,16.26)
    \Arc[dash,dashsize=10,clock](112,84)(32,-180,-360)
    \Arc[dash,dashsize=10,clock](112,60)(40,143.13,36.87)
    \Line(64,-44)(224,-44)
    \Arc[dash,dashsize=10](160,-24)(52,-157.38,-22.62)
    \Arc[dash,dashsize=10](160,-44)(48,-180,0)
    \Arc[dash,dashsize=10,clock](128,-44)(48,-180,-360)
    \Arc[dash,dashsize=10,clock](128,-64)(52,157.38,22.62)
    \Line(272,-44)(432,-44)
    \Arc[dash,dashsize=10](344,12)(25.298,-71.565,71.565)
    \Arc[dash,dashsize=10](360,12)(25.298,108.435,251.565)
    \Arc[dash,dashsize=10,clock](352,-29.6)(65.6,-167.32,-372.68)
    \Arc[dash,dashsize=10,clock](352,-92)(80,143.13,36.87)
    \Line(480,-44)(640,-44)
    \Arc[dash,dashsize=10](552,-4)(40.792,-78.69,78.69)
    \Arc[dash,dashsize=10](568,-4)(40.792,101.31,258.69)
    \Arc[dash,dashsize=10](560,4)(80,-143.13,-36.87)
    \Arc[dash,dashsize=10,clock](560,-29.6)(65.6,-167.32,-372.68)
    \Text(50,36)[lb]{\Large{\Black{$M_1$}}}
    \Text(200,36)[lb]{\Large{\Black{$M_2$}}}
    \Text(340,36)[lb]{\Large{\Black{$M_3$}}}
    \Text(50,-60)[lb]{\Large{\Black{$M_4$}}}
    \Text(200,-60)[lb]{\Large{\Black{$M_5$}}}
    \Text(340,-60)[lb]{\Large{\Black{$M_6$}}}
  \end{picture}
}
\end{center}

  \caption[]{\label{fig::MI2}
    Non-trivial master integrals contributing to the $n_l^2$ contribution.
    The same notation as in Fig.~\ref{fig::MI1} has been used.
  }
\end{figure}

Altogether we end up with 13 master integrals.  
Seven of them (shown in
Fig.~\ref{fig::MI1}) are products of one- and two-loop integrals
whereas the remaining six integrals (cf. Fig.~\ref{fig::MI2}) request a
dedicated investigation. We calculate them using the Dimensional Recurrence
and Analyticity (DRA) method introduced in~\cite{Lee:2009dh}.  In order to fix
the position and order of the poles of the integrals, we use
FIESTA~\cite{Smirnov:2008py,Smirnov:2009pb}. The remaining constants are 
fixed using the Mellin-Barnes
technique~\cite{Smirnov:1999gc,Tausk:1999vh,Smirnov:2013,Czakon:2005rk,Smirnov:2009up}. In
order to express the results in terms of the conventional multiple zeta values
we apply the PSLQ algorithm~\cite{Ferguson:1992} on high-precision numerical
results (with several hundreds of decimal digits).\footnote{Let us mention
  that the numerical evaluation of the factorizable four-loop master integrals
  for $a_l$ which reduce to the evaluation of the corresponding
  three-loop master integrals in higher orders of $\epsilon$ was undertaken in
  Ref.~\cite{Laporta:2001rc} as a warm-up before a future full four-loop
  calculation. This was done with the method of~\cite{Laporta:2001dd} based on
  difference equations. The achieved accuracy of several dozen of decimal
  digits was not enough for using PSLQ.}

The analytic results for the integrals in Fig.~\ref{fig::MI2} are listed in
Appendix~A. Results in terms of Gamma functions
for the integrals in Fig.~\ref{fig::MI1} are easily
obtained recursively using the formulae from the Appendix of
Ref.~\cite{Smirnov:2013}. For convenience also these results are given in
Appendix~A.

All results have been cross-checked numerically with the help
of {\tt FIESTA}~\cite{Smirnov:2009pb} where an accuracy of at least
four digits has been achieved.

%- }}}
%- {{{ Fermionic $n_l^2$ and $n_l^3$ contributions to $Z_2^{\rm OS}$ and $Z_m^{\rm OS}$:

\section{\label{sec::ZmZ2}Fermionic $n_l^2$ and $n_l^3$ contributions to $Z_m^{\rm OS}$ and $Z_2^{\rm OS}$}

Both $Z_m^{\rm OS}$ and $Z_2^{\rm OS}$ are obtained from the fermion two-point
functions $\Sigma(q)$ which can be cast in the form
\begin{eqnarray}
  \Sigma(q,m_q) &=&
  m_q\, \Sigma_1(q^2,m_q) + (\qsla - m_q)\, \Sigma_2(q^2,m_q)\,.
  \label{eq::defs::sigmadecomp}
\end{eqnarray}
Here $m_q$ represents a generic quark mass whereas bare, on-shell and
$\overline{\rm MS}$ quark masses are denoted by
$m_q^0$, $M_q$ and $\bar{m}_q$.

The derivation of ready-to-use formulae for $Z_m^{\rm OS}$ and $Z_2^{\rm OS}$ is
discussed at length in Refs.~\cite{Melnikov:2000qh,Marquard:2007uj}.
Thus, let us for 
convenience only repeat the final formulae which are applied in our
calculations. They read
\begin{eqnarray}
  Z_m^{\rm OS} &=& 1 + \Sigma_1(M_q^2,M_q)\,,
  \label{eq::defs::calcZm} \\
  \left( Z_2^{\rm OS} \right)^{-1} &=& 1 + 2M_q^2
  \frac{\partial}{\partial q^2} \Sigma_1(q^2,M_q) \Big|_{q^2 = M_q^2} +
  \Sigma_2(M_q^2,M_q) \,.
  \label{eq::defs::calcZ2}
\end{eqnarray}
The expressions on the right-hand side are computed by
introducing the momentum $Q$ with $Q^2 = M_q^2$ via
$q = Q(1+t)$ which leads to the equation
\begin{eqnarray}
  {\rm Tr} \left\{ \frac{\Qsla + M_q}{4M_q^2} \Sigma(q,M_q) \right\} &=&
  \Sigma_1(q^2,M_q) + t \Sigma_2(q^2,M_q) \nonumber \\
  &=& \Sigma_1(M_q^2,M_q)
  + \left( 2M_q^2 \frac{\partial}{\partial q^2} \Sigma_1(q^2,M_q)
  \Big|_{q^2 = M_q^2} \!\!+\! \Sigma_2(M_q^2,M_q) \right) t \nonumber\\ 
  &&  + {\cal O}(t^2) \,.
  \label{eq::defs::trace}
\end{eqnarray}
Hence, to obtain $Z_m^{\rm OS}$ one only needs to calculate 
$\Sigma_1$ for $q^2 = M_q^2$. 
To calculate $Z_2^{\rm OS}$, one has to compute the first derivative of
the self-energy diagrams. 
Note that the renormalization of the quark mass is taken into account
iteratively by explicitly calculating the corresponding 
counterterm diagrams. 

We write the perturbative expansion for $Z_m^{\rm OS}$ in terms of the
renormalized strong coupling as ($\gamma_E$ is the Euler-Mascheroni number)
\begin{eqnarray}
  Z_m^{\rm OS} &=& 
  1 + \frac{\alpha_s(\mu)}{\pi} \left(\frac{e^{\gamma_E}}{4 \pi}
  \right)^{-\epsilon} \delta Z_m^{(1)} 
  + \left(\frac{\alpha_s(\mu)}{\pi}\right)^2 \left(\frac{e^{\gamma_E}}{4 \pi}
  \right)^{-2\epsilon} \delta Z_m^{(2)} 
  \nonumber \\ &&\mbox{} 
  + \left(\frac{\alpha_s(\mu)}{\pi}\right)^3
  \left(\frac{e^{\gamma_E}}{4\pi} \right)^{-3\epsilon} \delta Z_m^{(3)}
  + \left(\frac{\alpha_s(\mu)}{\pi}\right)^4
  \left(\frac{e^{\gamma_E}}{4\pi} \right)^{-4\epsilon} \delta Z_m^{(4)}
  + \mathcal{O}\left(\alpha_s^5\right) 
  \,.
  \label{eq::mass::Zm}
\end{eqnarray}
This allows us to take the ratio between the on-shell and
$\overline{\rm
  MS}$~\cite{Chetyrkin:1997dh,Vermaseren:1997fq,Chetyrkin:2004mf} 
mass renormalization constant which is given by
\begin{eqnarray}
  z_m^{\rm OS}(\mu) 
  &=& \frac{\bar{m}_q(\mu)}{M_q} \,\,=\,\,\frac{Z_m^{\rm OS}}{Z_m^{\overline{\rm MS}}}
  \nonumber\\
  &=& 1 + \frac{\alpha_s(\mu)}{\pi}          \delta z_m^{(1)} 
  + \left(\frac{\alpha_s(\mu)}{\pi}\right)^2 \delta z_m^{(2)} 
  + \left(\frac{\alpha_s(\mu)}{\pi}\right)^3 \delta z_m^{(3)}
  + \left(\frac{\alpha_s(\mu)}{\pi}\right)^4 \delta z_m^{(4)}
  \nonumber \\ &&\mbox{} 
  + \mathcal{O}\left(\alpha_s^5\right) 
\end{eqnarray}
The coefficients $\delta z_m^{(i)}$  are by construction finite.

In the case of $Z_2^{\rm OS}$ we choose the bare coupling
as expansion parameter which in many applications turns out to be
convenient. Furthermore, the dependence on $\mu/M_q$ can be written in
factorized form which leads to shorter expressions.
Thus we have
\begin{eqnarray}
  Z_2^{\rm OS} &=& 
  1 + \frac{\alpha_s^0}{\pi} \left(\frac{e^{\gamma_E}}{4 \pi}
  \right)^{-\epsilon} \delta Z_2^{(1)} 
  + \left(\frac{\alpha_s^0}{\pi}\right)^2 \left(\frac{e^{\gamma_E}}{4 \pi}
  \right)^{-2\epsilon} \delta Z_2^{(2)} 
  \nonumber \\ &&\mbox{} 
  + \left(\frac{\alpha_s^0}{\pi}\right)^3
  \left(\frac{e^{\gamma_E}}{4\pi} \right)^{-3\epsilon} \delta Z_2^{(3)}
  + \left(\frac{\alpha_s^0}{\pi}\right)^4
  \left(\frac{e^{\gamma_E}}{4\pi} \right)^{-4\epsilon} \delta Z_2^{(4)}
  + \mathcal{O}\left(\left(\alpha_s^0\right)^5\right) 
  \,,
  \label{eq::Z2}
\end{eqnarray}
where each term $\delta Z_2^{(n)}$ contains a factor $(\mu^2/M_q^2)^{n\epsilon}$.

We refrain from repeating the one-, two- and three-loop results
for $Z_m^{\rm OS}$ and $Z_2^{\rm OS}$
since analytical expressions for general colour coefficients are available
in the literature~\cite{Melnikov:2000qh,Melnikov:2000zc,Marquard:2007uj}.
We split the four-loop coefficient according to the number of closed
massless fermion loops and write ($i\in\{m,2\}$)
\begin{eqnarray}
  \delta Z_i^{(4)} &=& 
  \delta Z_i^{(40)} + \delta Z_i^{(41)} n_l 
  + \delta Z_i^{(42)} n_l^2 + \delta Z_i^{(43)} n_l^3
  \,.
  \label{eq::delZnl}
\end{eqnarray}
with an analog notation for $\delta z_m^{(4)}$.

In the following we present analytical results for 
$\delta z_m^{(42)}$, $\delta z_m^{(43)}$, $\delta Z_2^{(42)}$
and $\delta Z_2^{(43)}$ which read
\newcommand{\lmm}{l_M}
\begin{eqnarray}
  \delta z_m^{(43)} &=&
  C_F T^3
  \left(
  \frac{\lmm^4}{144}
  + \frac{13 \lmm^3}{216}
  + \left(\frac{89}{432}
  + \frac{\pi ^2}{36}\right) \lmm^2
  + \lmm \left(\frac{\zeta_3}{3} + \frac{1301}{3888} + \frac{13 \pi ^2}{108}\right)
  \right.\nonumber\\&&\left.\mbox{}
  + \frac{317 \zeta_3}{432}
  + \frac{71 \pi ^4}{4320}
  + \frac{89 \pi    ^2}{648}
  + \frac{42979}{186624}\right)
  \,,
\end{eqnarray}			
\begin{eqnarray}			
  \delta z_m^{(42)} &=&
  C_F n_h T^3
  \left(
  \frac{\lmm^4}{48}+\frac{13\lmm^3}{72}+\frac{125
    \lmm^2}{144}+\frac{2489\lmm}{1296}
  +\frac{5\zeta_3}{144}-\frac{19\pi^4}{480}+\frac{\pi
    ^2}{6}+\frac{128515}{62208}
  \right)
  \nonumber\\&&\mbox{}
  + C_A C_F T^2
  \left(
  -\frac{11\lmm^4}{192}
  -\frac{91\lmm^3}{144}
  +
  \lmm^2
  \left(-\frac{1}{12}\pi^2a_1-\frac{\zeta_3}{4}
  -\frac{\pi^2}{8}-\frac{6539}{2304}\right)
  \right.\nonumber\\&&\left.\mbox{}
  +\lmm
  \left(\frac{a_1^4}{18}+\frac{1}{9}\pi^2
  a_1^2-\frac{11}{18}\pi^2a_1+\frac{4
    a_4}{3}-\frac{37\zeta_3}{16}-\frac{\pi
    ^4}{216}-\frac{29\pi
    ^2}{36}-\frac{15953}{2592}\right)
  \right.\nonumber\\&&\left.\mbox{}
  -\frac{1}{45}
  a_1^5+\frac{11
    a_1^4}{54}-\frac{2}{27}\pi^2
  a_1^3+\frac{11}{27}\pi
  ^2a_1^2-\frac{31\pi^4
    a_1}{1080}-\frac{103}{108}\pi
  ^2a_1+\frac{44a_4}{9}+\frac{8
    a_5}{3}
  \right.\nonumber\\&&\left.\mbox{}
  -\frac{41\zeta_5}{24}-\frac{13\pi^2\zeta_3}{48}-\frac{3245\zeta_3}{576}-\frac{4723\pi
    ^4}{51840}-\frac{527
    \pi^2}{384}-\frac{2708353}{497664}
  \right)
  \nonumber\\&&\mbox{}
  +C_F^2 T^2
  \left(
  +\frac{11\lmm^4}{384}
  +\frac{97\lmm^3}{576}
  +
  \lmm^2\left(\frac{1}{6}\pi^2
  a_1+\frac{\zeta_3}{8}-\frac{5\pi^2}{96}-\frac{157}{2304}
  \right)
  \right.\nonumber\\&&\left.\mbox{}
  +\lmm
  \left(-\frac{1}{9}a_1^4-\frac{2}{9}\pi^2
  a_1^2+\frac{11}{9}\pi^2a_1-\frac{8
    a_4}{3}-\frac{11\zeta_3}{8}+\frac{11\pi
    ^4}{216}-\frac{21\pi
    ^2}{32}-\frac{50131}{20736}
  \right)
  \right.\nonumber\\&&\left.\mbox{}
  +\frac{2
    a_1^5}{45}-\frac{11
    a_1^4}{27}+\frac{4}{27}\pi^2
  a_1^3-\frac{22}{27}\pi
  ^2a_1^2+\frac{31}{540}\pi^4
  a_1+\frac{103}{54}\pi^2
  a_1-\frac{88a_4}{9}-\frac{16
    a_5}{3}
  \right.\nonumber\\&&\left.\mbox{}
  +\frac{305\zeta_5}{48}+\frac{3\pi^2\zeta_3}{8}-\frac{2839\zeta_3}{576}+\frac{3683\pi^4}{51840}-\frac{5309\pi
    ^2}{3456}-\frac{2396921}{497664}\right)
  \,,
  \label{eq::zm42}
\end{eqnarray} 
\begin{eqnarray}
  \delta Z_2^{(43)}&=&
  C_F T^3 \left( \frac{\mu^2}{M_q^2} \right)^{4\epsilon} \left(
  \frac{1}{144\epsilon^4}
  +\frac{65}{864 \epsilon^3}
  +\frac{\frac{89}{192}+\frac{13\pi^2}{432}}{\epsilon^2}
  +\frac{\frac{151\zeta_3}{216}+\frac{73669}{31104}
    +\frac{845\pi^2}{2592}}{\epsilon}
  \right.\nonumber\\&&\left.\mbox{}  
  +\frac{9815\zeta_3}{1296}+\frac{589\pi^4}{4320}
  +\frac{1157\pi^2}{576}+\frac{2106347}{186624}
  \right)
  \,,
\end{eqnarray}
\begin{eqnarray}			
  \delta Z_2^{(42)} &=&
    \left( \frac{\mu^2}{M_q^2} \right)^{4\epsilon} \left[
    C_F n_h T^3
    \left(
    \frac{1}{36\epsilon^4}+\frac{187}{864 \epsilon^3}
    +\frac{\frac{10957}{5184}-\frac{5 \pi ^2}{108}}{\epsilon^2}
    \right.\right.\nonumber\\&&\left.\left.\mbox{}
    +\frac{\frac{2}{3} \pi ^2 a_1-\frac{71 \zeta_3}{54}
      -\frac{1013 \pi
    ^2}{2592}
  +\frac{349615}{31104}}{\epsilon}-\frac{10}{9}
    a_1^4-\frac{20}{9} \pi ^2 a_1^2
    +\frac{127}{18} \pi ^2
    a_1
    \right.\right.\nonumber\\&&\left.\left.\mbox{}
    -\frac{80 a_4}{3}
    -\frac{21719 \zeta_3}{1296}
    -\frac{\pi ^4}{360}
    -\frac{14027 \pi ^2}{15552}+\frac{13135057}{186624}\right)
    \right.\nonumber\\&&\left.\mbox{}
    +   C_A C_F T^2  \left(
    -\frac{11}{192\epsilon^4}-\frac{761}{1152 \epsilon^3}
      +\frac{-\frac{1}{6}
    \pi ^2 a_1+\frac{\zeta_3}{16}-\frac{13 \pi
    ^2}{192}-\frac{64433}{13824}}{\epsilon^2}
    \right.\right.\nonumber\\&&\left.\left.\mbox{}
    +\frac{\frac{5
    a_1^4}{18}+\frac{5}{9} \pi ^2 a_1^2-\frac{163}{72} \pi
    ^2 a_1+\frac{20 a_4}{3}+\frac{37 \zeta_3}{288}-\frac{647 \pi ^4}{8640}-\frac{1627 \pi
    ^2}{1152}-\frac{18287}{768}}{\epsilon}
    \right.\right.\nonumber\\&&\left.\left.\mbox{}
    -\frac{5}{9} a_1^5+\frac{815 a_1^4}{216}-\frac{50}{27} \pi ^2
    a_1^3+\frac{815}{108} \pi ^2 a_1^2+\frac{1}{18} \pi ^4
    a_1-\frac{281}{18} \pi ^2 a_1
    \right.\right.\nonumber\\&&\left.\left.\mbox{}
    +\frac{815 a_4}{9}+\frac{200 a_5}{3}
    -\frac{2079 \zeta_5}{32}-\frac{209 \pi ^2 \zeta_3}{144}-\frac{27977 \zeta_3}{1728}-\frac{7411 \pi ^4}{5760}
    \right.\right.\nonumber\\&&\left.\left.\mbox{}
    -\frac{436741 \pi
    ^2}{41472}-\frac{60973393}{497664}\right)
    \right.\nonumber\\&&\left.\mbox{}
    +C_F^2 T^2
    \left(
    \frac{11}{384 \epsilon^4}+\frac{47}{192 \epsilon^3}
      +\frac{\frac{1}{3} \pi ^2 a_1-\frac{5 \zeta_3}{16}-\frac{241 \pi
    ^2}{1152}+\frac{2363}{1536}}{\epsilon^2}
    \right.\right.\nonumber\\&&\left.\left.\mbox{}
    +\frac{-\frac{5}{9}
    a_1^4-\frac{10}{9} \pi ^2 a_1^2+\frac{163}{36} \pi ^2
    a_1-\frac{40 a_4}{3}-\frac{773 \zeta_3}{72}+\frac{383
    \pi ^4}{1728}-\frac{1181 \pi
    ^2}{576}+\frac{2893}{2304}}{\epsilon}
    \right.\right.\nonumber\\&&\left.\left.\mbox{}
    +\frac{10
    a_1^5}{9}-\frac{815 a_1^4}{108}+\frac{100}{27} \pi ^2
    a_1^3-\frac{815}{54} \pi ^2 a_1^2-\frac{1}{9} \pi ^4
    a_1+\frac{281}{9} \pi ^2 a_1
    \right.\right.\nonumber\\&&\left.\left.\mbox{}
    -\frac{1630
    a_4}{9}-\frac{400 a_5}{3}
    +\frac{7145 \zeta_5}{48}+\frac{187 \pi ^2 \zeta_3}{48}-\frac{50209 \zeta_3}{576}+\frac{8413 \pi ^4}{6480}
    \right.\right.\nonumber\\&&\left.\left.\mbox{}
    -\frac{75089 \pi
    ^2}{4608}-\frac{261181}{55296}\right)
\right]
  \label{eq::Z242}
  \,,
\end{eqnarray}
where $\lmm=\ln\mu^2/M_q^2$, $\zeta_n$ is Riemann's zeta function, $a_1=\ln2$
and $a_n=\mbox{Li}_n(1/2)$ ($n\ge1$).  In the case of QCD the colour factors
take the values $C_A=3, C_F=4/3$ and $T=1/2$.  In Eqs.~(\ref{eq::zm42})
and~(\ref{eq::Z242}) the contributions from closed heavy quark loops are
marked by $n_h=1$ which has been introduced for illustration.

In order to get an impression of the numerical size of the newly calculated
terms we evaluate $z_m^{\rm OS}$ for $\mu=M_q$. After inserting the numerical values
for the colour factors we obtain ($A_s\equiv \alpha_s(M_q)/\pi$)
\begin{eqnarray}
  z_m^{\rm OS} &=& 1 - A_s 1.333 
  + A_s^2 \left(-14.229 - 0.104\,n_h + 1.041\,n_l\right) 
  \nonumber\\&&\mbox{}
  + A_s^3 \left(-197.816 - 0.827\,n_h - 0.064\,n_h^2 + 26.946\,n_l 
  - 0.022\,n_h n_l - 0.653\,n_l^2\right)
  \nonumber\\&&\mbox{}
 + A_s^4 \left(-43.465\,n_l^2 - 0.017\,n_h n_l^2 + 0.678\,n_l^3 + \ldots\right)
 + {\cal O}\left(A_s^5\right)
 \,,
 \label{eq::zm}
\end{eqnarray}
where the ellipses indicate $n_l$ independent contributions and terms
proportional to $n_l$ which have not been computed.
One observes that the $n_l^2$ contribution at two loops and the $n_l^3$
contribution at three loops are quite small. This is in contrast to the 
linear $n_l$ terms which can become quite 
sizeable. E.g., setting $n_l=5$, which corresponds to the case of the top
quark, we obtain (for $n_h=1$)
\begin{eqnarray}
  z_m^{\rm OS} &=& 1 - A_s 1.333 
  + A_s^2 \left(-14.332 + 5.207_{n_l}\right) 
  \nonumber\\&&\mbox{}
  + A_s^3 \left(-198.707 + 134.619_{n_l} - 16.317_{n_l^2} \right)
  \nonumber\\&&\mbox{}
 + A_s^4 \left(-1087.060_{n_l^2} + 84.768_{n_l^3} + \ldots\right)
 + {\cal O}\left(A_s^5\right)
 \,.
 \label{eq::zm5}
\end{eqnarray}
At two-loop order the $n_l$ contribution is only a factor of three smaller
than the $n_l$-independent term, however, with an opposite sign.  At three
loops the linear-$n_l$ term has almost the same order of magnitude than the
constant contribution but again a different sign. It is remarkable that for
$n_l=5$ the coefficient of the four-loop $n_l^2$ term is more than a factor of
five larger than the $n_l$-independent term at order $\alpha_s^3$.

Let us finally compare our results with the approximate expressions
obtained in Ref.~\cite{Beneke:1994qe} in the large-$\beta_0$ approximation.
In Ref.~\cite{Beneke:1994qe} one finds for the quantity
$M_q/\bar{m}_q(\bar{m}_q)$ the result ($a_s\equiv \alpha_s(\bar{m}_q)/\pi$)
\begin{eqnarray}
  \frac{M_q}{\bar{m}_q(\bar{m}_q)}\Bigg|_{\mbox{large}-\beta_0} &=& 
  1 + a_s 1.333 + a_s^2 \left( 17.186 - 1.041 n_l \right)
  \nonumber\\&&\mbox{}
  + a_s^3 \left( 177.695 -  21.539 n_l + 0.653 n_l^2 \right)
  \nonumber\\&&\mbox{}
  + a_s^4 \left( 3046.294 - 553.872 n_l + 33.568 n_l^2 - 0.678 n_l^3 \right)
  \,,
  \label{eq::izm_largeb0}
\end{eqnarray}
where for the renormalization scale $\mu=\bar{m}_q$ has been chosen.
The coefficients of Eq.~(\ref{eq::izm_largeb0}) should be compared with
our findings which read
\begin{eqnarray}
  \frac{M_q}{\bar{m}_q(\bar{m}_q)} &=& 
  1 + a_s 1.333 + a_s^2 \left( 13.443 - 1.041 n_l \right)
  \nonumber\\&&\mbox{}
  + a_s^3 \left( 190.595 -  26.655 n_l + 0.653 n_l^2 \right)
  \nonumber\\&&\mbox{}
  + a_s^4 \left( c_0 + c_1 n_l  + 43.396 n_l^2 - 0.678 n_l^3 \right)
  \,,
  \label{eq::izm}
\end{eqnarray}
where $c_0$ and $c_1$ are not yet known.
By construction one finds agreement for the coefficient of $n_l^3$
since it has been used as input in Ref.~\cite{Beneke:1994qe}.
As far as the $n_l^2$ term is concerned the exact coefficient is
predicted with an accuracy of about 30\%.

%- }}}
%- {{{ Fermionic $n_l^2$ and $n_l^3$ contributions to $a_\mu$:

\section{\label{sec::gm2}Fermionic $n_l^2$ and $n_l^3$ contributions to $a_\mu$}

It is convenient to introduce the form factors
$F_1$ and $F_2$ of the photon-lepton vertex as
\begin{eqnarray}
  \Gamma^\mu(q,p) = F_1(q^2) \gamma^\mu 
  + i \frac{ F_2(q^2) }{2M_l} \sigma_{\mu\nu} q^\nu
  \,,
  \label{eq::Gamma}
\end{eqnarray}
where $q$ is the incoming momentum in the photon line and $M_l$ is the lepton 
mass. The anomalous magnetic moment is given by
\begin{eqnarray}
  a_l &=& \left(\frac{g-2}{2}\right)_l \,\,=\,\, F_2(0)
  \,.
\end{eqnarray}
In Eq.~(\ref{eq::Gamma}) also the momentum $p=(p_1+p_2)/2$ has been 
introduced where $p_1^2=p_2^2=M_l^2$ are the momenta flowing through the
external fermion lines (see Fig.~\ref{fig::diags_gm2} for the directions
of the momenta).

The evaluation of $a_l$ requires that $\Gamma^\mu(q,p)$ is computed in the
limit $q\to0$.
Due to the factor $q^\nu$ in front of $F_2$ in Eq.~(\ref{eq::Gamma})
one has to perform an expansion of $\Gamma^\mu(q,p)$ up to linear terms in 
$q$ which can be written as
\begin{eqnarray}
  \Gamma^\mu(q,p) &=& X^\mu(p) + q_\nu Y^{\mu\nu}(p) + {\cal O}\left(q^2\right)
  \,,
\end{eqnarray}
with $p^2=M_l^2$.
$F_2$ is conveniently obtained after the application of a projector
given by (see, e.g., Ref.~\cite{KrausePhD})
\begin{eqnarray}
  a_l &=& \frac{1}{2 M_l^2 (D-1) (D-2)}\mbox{Tr}\Bigg[ 
    \frac{D-2}{2}\left(M_l^2\gamma_\mu-Dp_\mu
    p\!\!\!/-(D-1)M_lp_\mu\right)X^\mu
    \nonumber\\&&\mbox{}
    +\frac{M_l}{4}
    \left( p\!\!\!/ + M_l \right)
    \left[ \gamma_\nu , \gamma_\mu \right]
    \left( p\!\!\!/ + M_l \right)
    Y^{\mu\nu}
    \Bigg]
  \,,
\end{eqnarray}
and thus $a_l$ is reduced to the evaluation of 
on-shell two-point functions
as described in Section~\ref{sec::os-ints}. 

We define the loop expansion of $a_l$ in analogy to Eq.~(\ref{eq::mass::Zm})
(with $\alpha_s$ replaced by the fine structure constant)
and introduce the same splitting according to the number of massless
lepton loops as in Eq.~(\ref{eq::delZnl}).

The Feynman diagrams contributing to $a_l$ respectively the coefficients of
$\alpha^n$ and $n_l^k$ can be subdivided to two classes: (i) the one where the
external photon couples to the lepton at hand and (ii) the one
where it couples to a lepton present in a closed loop. Sample diagrams 
are given in Fig.~\ref{fig::diags_gm2}. 
In the following we refer to the diagrams of class (ii) as ``light-by-light''
contribution in analogy to the corresponding hadronic part.

In this paper four-loop corrections contributing to class~(i) are evaluated
which contain two or three closed massless fermion loops. They are used in
order to compute electron loop contributions to $a_\mu$ neglecting 
terms of order $M_e/M_\mu$.

For the diagrams in class~(i) we can proceed as follows: In a first step we
renormalize the fine structure constant in the $\overline{\rm MS}$ scheme,
$\bar\alpha(\mu)$. The corresponding renormalization constant is easily
obtained from the one for $\alpha_s$ after specifying the colour factors to
QED. The $\overline{\rm MS}$ scheme has the advantage that the electron mass
can be set to zero (which is not the case for the diagrams in class~(ii)).
After renormalizing the muon mass in the on-shell scheme
we obtain a finite expression for $a_\mu$
which shows an explicit dependence on $\ln(\mu^2/M_\mu^2)$.

In a next step we replace $\bar\alpha(\mu)$ by its on-shell counterpart using
the corresponding relation up to three loops. It can best be calculated by
considering the photon two point function
\begin{eqnarray}
  (q^2 g^{\mu\nu} - q^\mu q^\nu)\Pi(q^2) &=& i \int \mathrm{d}x \,e^{i q x }
  \langle 0 | j^\mu(x) j^{\nu}(x) | 0 \rangle
\end{eqnarray}
and employing the on-shell renormalization condition $\Pi(q^2 = 0) = 0$.  The
form of the renormalization condition reduces the problem to the calculation
of two-scale vacuum integrals at three loops. Note, that for the
renormalization of the fermion masses in the on-shell scheme the dependence on
both masses has to be taken into account.
In the limit $M_e \ll M_\mu$ we obtain (see also
Refs.~\cite{Broadhurst:1991fi,Broadhurst:1992za,Baikov:2012rr}) 
\newcommand{\lone}{L_\mu}
\newcommand{\ltwo}{L_e}
\begin{eqnarray}
  \frac{\bar\alpha(\mu)}{\alpha} &=&
  1 
  + \frac{\alpha}{3\pi}\left( \lone + \ltwo \right)
  + \left(\frac{\alpha}{\pi}\right)^2 \left[ 
    \frac{15}{8} + \frac{ \lone + \ltwo }{4} 
    + \frac{\left( \lone + \ltwo \right)^2}{9}
    \right]
  \nonumber\\&&\mbox{}
  + \left(\frac{\alpha}{\pi}\right)^3 \left(
    \frac{\ltwo^3}{27}
    + \frac{\lone\ltwo^2}{9}
    + \frac{5\ltwo^2}{24}
    + \frac{79\ltwo}{144}
    - \frac{695}{648} 
    + \frac{\pi^2}{9}
    + \frac{7\zeta_3}{64}
    + \ldots \right)
  + {\cal O}(\alpha_s^4)
  \nonumber\\
  \label{eq::alMSalOS}
\end{eqnarray}
with $\lone=\ln(\mu^2/M_\mu^2)$ and $\ltwo=\ln(\mu^2/M_e^2)$.
The ellipses in the coefficient of $(\alpha/\pi)^3$
indicate terms which we left out since they are irrelevant 
for the $n_l^2$ contribution discussed in this paper. The complete result 
containing the exact dependence on $M_e/M_\mu$ is presented in Appendix~B.
Note that the result in Eq.~(\ref{eq::alMSalOS}) can be obtained from the one
provided in Ref.~\cite{Baikov:2012rr} where the relation between
$\bar\alpha(\mu)$ and $\alpha$ is given for one massive lepton.

Also in the case of $a_l$ we refrain from listing the lower-order results
which can be found in the
literature~\cite{Laporta:1996mq,Jegerlehner:2008zza,Jegerlehner:2009ry,Aoyama:2012wk,Aoyama:2012wj}.
Rather we concentrate on the new correction terms at four loops.  Adopting the
notation from Eq.~(\ref{eq::delZnl}) we obtain the following results for the
$n_l^3$ contribution
\newcommand{\Mu}{M_\mu} \newcommand{\Me}{M_e}
\newcommand{\lnue}{L_{\mu e}}
\begin{eqnarray}
  a_\mu^{(43)} &=&
   \frac{1}{54} \lnue^3
   -\frac{25}{108} \lnue^2+\left(\frac{317}{324}+\frac
    {\pi ^2}{27}\right)
    \lnue -\frac{2 \zeta_3}{9}
    -\frac{25 \pi ^2}{162}-\frac{8609}{5832}
  \nonumber\\
  &\approx& 7.196\,66\,,
  \label{eq::amu43}
\end{eqnarray}
where $\lnue=\ln(\Mu^2/\Me^2)$.
The approximate results have been obtained with the help of~\cite{Mohr:2012tt}
$M_\mu/M_e = 206.768 2843 (52)$.
The result in Eq.~(\ref{eq::amu43}) 
agrees with the one in Ref.~\cite{Laporta:1993ds,Aguilar:2008qj}.

In the case of the $n_l^2$ contribution we split
$a_\mu^{(42)}$ into two parts. The first one ($a_\mu^{(42)a}$) 
corresponds to the 
diagrams containing two closed fermion loops and the second one
($a_\mu^{(42)b}$) originates from diagrams with three closed fermion loops
where one of them is a muon and two are electron loops.
Thus, we have
\begin{eqnarray}
  a_\mu^{(42)} &=& a_\mu^{(42)a} + a_\mu^{(42)b}\,,
  \nonumber
\end{eqnarray}
with
\begin{eqnarray}
  a_\mu^{(42)a} &=&
    \lnue^2
    \left[\pi ^2
      \left(\frac{5}{36}-\frac{a_1}{6}\right)+\frac{\zeta_3}{4}
      -\frac{13}{24}\right]  
   + \lnue \left[-\frac{a_1^4}{9}+\pi ^2
    \left(-\frac{2 a_1^2}{9}+\frac{5
    a_1}{3}-\frac{79}{54}\right)
    \right.\nonumber\\&&\left.\mbox{}
    -\frac{8 a_4}{3}-3 \zeta_3+\frac{11 \pi ^4}{216}
    +\frac{23}{6}\right]
   -\frac{2 a_1^5}{45}+\frac{5 a_1^4}{9}
    +\pi ^2 \left(-\frac{4
    a_1^3}{27}+\frac{10 a_1^2}{9}
    \right.\nonumber\\&&\left.\mbox{}
    -\frac{235
    a_1}{54}-\frac{\zeta_3}{8}+\frac{595}{162}\right)
    +\pi ^4 \left(-\frac{31
    a_1}{540}-\frac{403}{3240}\right)+\frac{40 a_4}{3}+\frac{16
    a_5}{3}-\frac{37 \zeta_5}{6}
    \nonumber\\&&\mbox{}
    +\frac{11167 \zeta_3}{1152}-\frac{6833}{864}
  \nonumber\\
  &\approx&
  -3.624\,27\,,
  \label{eq::amu42a}
  \\
  a_\mu^{(42)b} &=&
  \left(\frac{119}{108}-\frac{\pi ^2}{9}\right) \lnue^2+\left(\frac{\pi
    ^2}{27}-\frac{61}{162}\right) \lnue-\frac{4 \pi ^4}{45}+\frac{13 \pi
    ^2}{27}+\frac{7627}{1944}
  \nonumber\\
  &\approx&
  0.494\,05
  \,.
\end{eqnarray}
$a_\mu^{(42)b}$ agrees with Ref.~\cite{Laporta:1993ds,Aguilar:2008qj}.
Analytical results for $a_\mu^{(42)a}$ are not present in the literature since
corrections originating from diagrams as the third one in the first row of
Fig.~\ref{fig::diags_gm2} have not been considered yet.  However, we can
perform a numerical comparison with the results from
Refs.~\cite{Kinoshita:2004wi,Aoyama:2012wk}\footnote{In in
  Ref.~\cite{Aoyama:2012wk} the contributions from closed electron and muon
  loops are always added whereas in our result at least two closed electron
  loops are present.  We are deeply grateful to the authors of
  Ref.~\cite{Aoyama:2012wk} for providing us the results for the contributions
  containing only electron loops Eq.~(\ref{eq::amunum}).}
which reads
\begin{eqnarray} 
  a_\mu^{(42)a}\Big|_{\rm num} &=& -3.642\,04(1\,12)\,,
  \label{eq::amunum}
\end{eqnarray}
There is a good agreement with the analytic result in 
Eq.~(\ref{eq::amu42a}). The deviation can be explained by corrections of
order $M_e/M_\mu \approx 0.005$ or $(M_e/M_\mu)^2 \ln^3 M_\mu/M_e \approx
0.004$~\cite{Laporta:1993ds,Aguilar:2008qj} which are absent in our analytic
expressions.

%- }}}
%- {{{ Conclusions:

\section{\label{sec::concl}Conclusions}

In this paper the first steps towards the evaluation of four-loop on-shell
integrals have been undertaken. 
As an application within QCD we have computed the
contributions involving two massless quark loops to the on-shell
renormalization constants $Z_2^{\rm OS}$ and $Z_m^{\rm OS}$. As an application in QED
we have considered the contribution from four-loop diagrams involving two or
three closed electron loops to the anomalous magnetic moment of the
muon excluding, however, the light-by-light contribution.

We describe in some detail the techniques and the programs which have been
used for the calculation. 
We are confident that they are generic enough to be applied to the
$n_l^1$ and non-fermionic contribution. The only bottleneck might be the
analytic evaluation of the master integrals so that maybe numerical methods
have to be applied.

%- }}}

%- {{{ Ackn.:

\section*{Acknowledgements}

We would like to thank M.~Nio for useful communications concerning the
numerical results for $a_\mu$. We also thank K.G.~Chetyrkin for carefully
reading the manuscript.
This work was supported by the DFG through the SFB/TR~9 ``Computational
Particle Physics''.
The work of R.L, A.S. and V.S. was also supported by the Russian
Foundation for Basic Research through grant 11-02-01196.
The Feynman diagrams were drawn with {\tt
  JaxoDraw}~\cite{Vermaseren:1994je,Binosi:2008ig}.

%- }}}

%- {{{ Appendix:

\begin{appendix}

%- {{{ A: MIs

\section*{\label{app::MIs}Appendix A: Analytic results for the master integrals}

In this appendix we provide the analytic results of all master integrals
where we assume an integration measure ${\rm d}^D k/(i\pi)^{D/2}$ with
$D=4-2\epsilon$. Furthermore we write scalar propagators of particles with
mass $M$ in the form $1/(-k^2+M^2)$. For convenience we set $M=1$
in the final result since the dependence on $M$ can easily be restored.

The analytic results for the integrals in Fig.~\ref{fig::MI1} read
\begin{align}
  L_1 &=\frac{\Gamma \left(5-\frac{3 D}{2}\right) \Gamma
    \left(1-\frac{D}{2}\right) \Gamma \left(2-\frac{D}{2}\right)^2 \Gamma
    \left(\frac{D}{2}-1\right)^4 \Gamma (3 D-9)}{\Gamma (D-2)^2 \Gamma (2
    D-5)}
  \,,\nonumber\\
  L_2 &=\frac{\Gamma (3-D)^2 \Gamma \left(2-\frac{D}{2}\right)^2 \Gamma
    \left(\frac{D}{2}-1\right)^4 \Gamma (2 D-5)^2}{\Gamma (D-2)^2 \Gamma
    \left(\frac{3 D}{2}-3\right)^2}
  \,,\\
  L_3 &=\frac{\Gamma (5-2 D) \Gamma \left(4-\frac{3 D}{2}\right) \Gamma
    \left(\frac{D}{2}-1\right)^4 \Gamma (4 D-9)}{\Gamma (2 D-4) \Gamma
    \left(\frac{5 D}{2}-5\right)}
  \,,\\
  L_4 &=\frac{\Gamma (6-2 D) \Gamma \left(5-\frac{3 D}{2}\right) \Gamma
    \left(2-\frac{D}{2}\right)^2 \Gamma \left(\frac{D}{2}-1\right)^5 \Gamma
    \left(\frac{3 D}{2}-4\right) \Gamma (4 D-11)}{\Gamma (4-D) \Gamma (D-2)^2
    \Gamma (2 D-5) \Gamma \left(\frac{5 D}{2}-6\right)}
  \,,\\
  L_5 &=\frac{\Gamma (6-2 D) \Gamma (3-D) \Gamma \left(2-\frac{D}{2}\right)
    \Gamma \left(\frac{D}{2}-1\right)^5 \Gamma (4 D-11)}{\Gamma (D-2) \Gamma
    \left(\frac{3 D}{2}-3\right) \Gamma \left(\frac{5 D}{2}-6\right)}
  \,,\\
  L_6 &=\frac{\Gamma (7-2 D) \Gamma \left(2-\frac{D}{2}\right)^3 \Gamma
    \left(\frac{D}{2}-1\right)^6 \Gamma (4 D-13)}{\Gamma (D-2)^3 \Gamma
    \left(\frac{5 D}{2}-7\right)}
  \,,\\
  L_7 &=\frac{\Gamma (6-2 D) \Gamma \left(5-\frac{3 D}{2}\right)^2 \Gamma
    \left(2-\frac{D}{2}\right)^2 \Gamma \left(\frac{D}{2}-1\right)^4 \Gamma
    \left(\frac{3 D}{2}-4\right)}{\Gamma (10-3 D) \Gamma (D-2)^2 \Gamma
    \left(\frac{D}{2}\right)}
  \,.
\end{align}

The analytic results for the integrals in Fig.~\ref{fig::MI2} read
\begin{align}
  e^{4\epsilon\gamma_E}M_1 &=
\frac{5}{24 \epsilon ^4}+\frac{25}{24 \epsilon ^3}+
\biggl(\frac{205}{96}+\frac{17 \pi ^2}{72}\biggr) \epsilon ^{-2}+
\biggl(-\frac{323}{96}+\frac{85 \pi ^2}{72}+\frac{79 \zeta _3}{18}\biggr) \epsilon ^{-1}+
\biggl(-\frac{55241}{1152}
\nonumber\\
&+\frac{409 \pi ^2}{288}+\frac{395 \zeta _3}{18}+\frac{\pi ^4}{8}\biggr) \epsilon ^0-
\biggl(\frac{733351}{3456}+\frac{4199 \pi ^2}{288}-\frac{1943 \zeta _3}{72}-\frac{5 \pi ^4}{8}-\frac{499 \pi ^2 \zeta _3}{54}
\nonumber\\
&-\frac{407 \zeta _5}{6}\biggr) \epsilon ^1-
\biggl(\frac{14346449}{41472}+\frac{383045 \pi ^2}{3456}+\frac{19057 \zeta _3}{72}+\frac{437 \pi ^4}{96}-\frac{2495 \pi ^2 \zeta _3}{54}-\frac{2035 \zeta _5}{6}
\nonumber\\
&+\frac{2027 \pi ^6}{11340}-\frac{2285 \zeta _3^2}{27}\biggr) \epsilon ^2-
\biggl(-\frac{391938053}{124416}+\frac{3517963 \pi ^2}{10368}+\frac{1751323 \zeta _3}{864}+\frac{93347 \pi ^4}{1440}
\nonumber\\
&-\frac{443 \pi ^2 \zeta _3}{216}+\frac{21905 \zeta _5}{24}+\frac{2027 \pi ^6}{2268}-\frac{11425 \zeta _3^2}{27}-\frac{2477 \pi ^4 \zeta _3}{90}-\frac{9223 \pi ^2 \zeta _5}{90}+\frac{11681 \zeta _7}{42}\biggr) \epsilon ^3
\nonumber\\
&+
O\left(\epsilon ^4\right)
  \,,
\end{align}
\begin{align}
 e^{4\epsilon\gamma_E} M_2 &=
-\frac{5}{12 \epsilon ^4}-\frac{23}{8 \epsilon ^3}-
\biggl(\frac{433}{48}+\frac{29 \pi ^2}{36}\biggr) \epsilon ^{-2}-
\biggl(-\frac{297}{32}+\frac{191 \pi ^2}{24}+\frac{275 \zeta _3}{18}\biggr) \epsilon ^{-1}-
\biggl(-\frac{22765}{64}
\nonumber\\
&+\frac{7177 \pi ^2}{144}-24 \pi ^2 a_1+\frac{2273 \zeta _3}{12}+\frac{125 \pi ^4}{72}\biggr) \epsilon ^0-
\biggl(-\frac{411105}{128}+\frac{8085 \pi ^2}{32}-324 \pi ^2 a_1
\nonumber\\
&+\frac{105463 \zeta _3}{72}+\frac{2747 \pi ^4}{240}+80 \pi ^2 a_1^2+40 a_1^4+\frac{1595 \pi ^2 \zeta _3}{54}+\frac{3223 \zeta _5}{6}+960 a_4\biggr) \epsilon ^1
\nonumber\\
&-
\biggl(-\frac{16944559}{768}+\frac{216731 \pi ^2}{192}-2706 \pi ^2 a_1+\frac{146091 \zeta _3}{16}+\frac{43757 \pi ^4}{1440}+1080 \pi ^2 a_1^2
\nonumber\\
&+540 a_1^4+\frac{16 \pi ^4 a_1}{3}-\frac{800}{3} \pi ^2 a_1^3-80 a_1^5+\frac{9785 \pi ^2 \zeta _3}{36}-\frac{52351 \zeta _5}{20}+\frac{112339 \pi ^6}{22680}+\frac{15125 \zeta _3^2}{54}
\nonumber\\
&+12960 a_4+9600 a_5\biggr) \epsilon ^2-
\biggl(-\frac{68697721}{512}+\frac{589805 \pi ^2}{128}-18165 \pi ^2 a_1+\frac{4851365 \zeta _3}{96}
\nonumber\\
&-\frac{94853 \pi ^4}{960}+9020 \pi ^2 a_1^2+4510 a_1^4+72 \pi ^4 a_1-3600 \pi ^2 a_1^3-1080 a_1^5+\frac{336415 \pi ^2 \zeta _3}{216}
\nonumber\\
&-\frac{8849321 \zeta _5}{120}+32640 s_6-\frac{351599 \pi ^6}{15120}-104 \pi ^4 a_1^2+744 \pi ^2 a_1^4+\frac{400 a_1^6}{3}+944 \pi ^2 a_1 \zeta _3
\nonumber\\
&-\frac{375097 \zeta _3^2}{36}+\frac{6875 \pi ^4 \zeta _3}{108}+\frac{93467 \pi ^2 \zeta _5}{90}+\frac{652775 \zeta _7}{42}+108240 a_4+129600 a_5+96000 a_6
\nonumber\\
&+1856 \pi ^2 a_4\biggr) \epsilon ^3+
O\left(\epsilon ^4\right)
  \,,
\end{align}
\begin{align}
 e^{4\epsilon\gamma_E} M_3 &=
-\frac{1}{6 \epsilon ^4}-\frac{7}{6 \epsilon ^3}-
\biggl(\frac{10}{3}+\frac{13 \pi ^2}{18}\biggr) \epsilon ^{-2}-
\biggl(-\frac{61}{6}+\frac{73 \pi ^2}{18}+\frac{118 \zeta _3}{9}\biggr) \epsilon ^{-1}-
\biggl(-\frac{851}{4}+\frac{83 \pi ^2}{18}
\nonumber\\
&+\frac{637 \zeta _3}{9}+\frac{37 \pi ^4}{10}\biggr) \epsilon ^0-
\biggl(-\frac{14861}{8}-\frac{3467 \pi ^2}{36}+\frac{1003 \zeta _3}{18}+\frac{1121 \pi ^4}{60}+\frac{1894 \pi ^2 \zeta _3}{27}
\nonumber\\
&+\frac{16018 \zeta _5}{15}\biggr) \epsilon ^1-
\biggl(-\frac{613975}{48}-\frac{25981 \pi ^2}{24}-\frac{68293 \zeta _3}{36}+\frac{83 \pi ^4}{24}+\frac{9559 \pi ^2 \zeta _3}{27}+\frac{79891 \zeta _5}{15}
\nonumber\\
&+\frac{59501 \pi ^6}{2835}+\frac{17704 \zeta _3^2}{27}\biggr) \epsilon ^2-
\biggl(-\frac{7539347}{96}-\frac{382349 \pi ^2}{48}-\frac{482627 \zeta _3}{24}-\frac{426659 \pi ^4}{720}
\nonumber\\
&+\frac{3757 \pi ^2 \zeta _3}{54}+\frac{2525 \zeta _5}{6}+\frac{43201 \pi ^6}{420}+\frac{88585 \zeta _3^2}{27}+\frac{17204 \pi ^4 \zeta _3}{45}+\frac{206434 \pi ^2 \zeta _5}{45}
\nonumber\\
&+\frac{1267243 \zeta _7}{21}\biggr) \epsilon ^3+
O\left(\epsilon ^4\right)
  \,,
\end{align}
\begin{align}
 e^{4\epsilon\gamma_E} M_4 &=
-\frac{1}{3 \epsilon ^4}-\frac{5}{2 \epsilon ^3}-
\biggl(\frac{55}{6}+\frac{4 \pi ^2}{9}\biggr) \epsilon ^{-2}-
\biggl(3+\frac{19 \pi ^2}{3}+\frac{56 \zeta _3}{9}\biggr) \epsilon ^{-1}-
\biggl(-250+\frac{1925 \pi ^2}{36}
\nonumber\\
&-32 \pi ^2 a_1+\frac{464 \zeta _3}{3}+\frac{19 \pi ^4}{45}\biggr) \epsilon ^0-
\biggl(-\frac{5091}{2}+\frac{2811 \pi ^2}{8}-432 \pi ^2 a_1+\frac{14797 \zeta _3}{9}+\frac{17 \pi ^4}{90}
\nonumber\\
&+\frac{320}{3} \pi ^2 a_1^2+\frac{160 a_1^4}{3}+\frac{332 \pi ^2 \zeta _3}{27}+\frac{1556 \zeta _5}{15}+1280 a_4\biggr) \epsilon ^1-
\biggl(-\frac{55049}{3}+\frac{95693 \pi ^2}{48}
\nonumber\\
&-3608 \pi ^2 a_1+\frac{24831 \zeta _3}{2}-\frac{1084 \pi ^4}{45}+1440 \pi ^2 a_1^2+720 a_1^4+\frac{160 \pi ^4 a_1}{9}-\frac{3200}{9} \pi ^2 a_1^3
\nonumber\\
&-\frac{320 a_1^5}{3}+\frac{1046 \pi ^2 \zeta _3}{9}-8246 \zeta _5+\frac{772 \pi ^6}{2835}+\frac{2648 \zeta _3^2}{27}+17280 a_4+12800 a_5\biggr) \epsilon ^2
\nonumber\\
&-
\biggl(-\frac{458141}{4}+\frac{329467 \pi ^2}{32}-24220 \pi ^2 a_1+\frac{938425 \zeta _3}{12}-\frac{9979 \pi ^4}{36}+\frac{36080}{3} \pi ^2 a_1^2
\nonumber\\
&+\frac{18040 a_1^4}{3}+240 \pi ^4 a_1-4800 \pi ^2 a_1^3-1440 a_1^5+\frac{19930 \pi ^2 \zeta _3}{27}-\frac{353044 \zeta _5}{3}+44800 s_6
\nonumber\\
&-\frac{76904 \pi ^6}{945}-176 \pi ^4 a_1^2+976 \pi ^2 a_1^4+\frac{1600 a_1^6}{9}+\frac{4160}{3} \pi ^2 a_1 \zeta _3-\frac{155668 \zeta _3^2}{9}+\frac{4304 \pi ^4 \zeta _3}{135}
\nonumber\\
&+\frac{7304 \pi ^2 \zeta _5}{45}+\frac{4616 \zeta _7}{21}+144320 a_4+172800 a_5+128000 a_6+\frac{6272 \pi ^2 a_4}{3}\biggr) \epsilon ^3
\nonumber\\
&+
O\left(\epsilon ^4\right)
  \,,
\end{align}
\begin{align}
e^{4\epsilon\gamma_E}  M_5 &=
-\frac{1}{12 \epsilon ^4}-\frac{13}{24 \epsilon ^3}-
\biggl(\frac{15}{16}+\frac{13 \pi ^2}{36}\biggr) \epsilon ^{-2}-
\biggl(-\frac{1135}{96}+\frac{169 \pi ^2}{72}+\frac{86 \zeta _3}{9}\biggr) \epsilon ^{-1}-
\biggl(-\frac{28699}{192}
\nonumber\\
&+\frac{65 \pi ^2}{16}+\frac{559 \zeta _3}{9}+\frac{149 \pi ^4}{90}\biggr) \epsilon ^0-
\biggl(-\frac{144429}{128}-\frac{14755 \pi ^2}{288}+\frac{227 \zeta _3}{2}+\frac{1937 \pi ^4}{180}
\nonumber\\
&+\frac{1118 \pi ^2 \zeta _3}{27}+\frac{7604 \zeta _5}{15}\biggr) \epsilon ^1-
\biggl(-\frac{5327075}{768}-\frac{373087 \pi ^2}{576}-\frac{45889 \zeta _3}{36}+\frac{749 \pi ^4}{40}
\nonumber\\
&+\frac{7267 \pi ^2 \zeta _3}{27}+\frac{49426 \zeta _5}{15}+\frac{14053 \pi ^6}{1620}+\frac{14063 \zeta _3^2}{27}\biggr) \epsilon ^2-
\biggl(-\frac{58275695}{1536}-\frac{625859 \pi ^2}{128}
\nonumber\\
&-\frac{1192693 \zeta _3}{72}-\frac{168143 \pi ^4}{720}+\frac{2951 \pi ^2 \zeta _3}{6}+5805 \zeta _5+\frac{182689 \pi ^6}{3240}+\frac{182819 \zeta _3^2}{54}
\nonumber\\
&+\frac{51013 \pi ^4 \zeta _3}{270}+\frac{98852 \pi ^2 \zeta _5}{45}+\frac{1021711 \zeta _7}{42}\biggr) \epsilon ^3+
O\left(\epsilon ^4\right)
  \,,
\end{align}
\begin{align}
e^{4\epsilon\gamma_E}  M_6 &=
-\frac{1}{12 \epsilon ^4}-\frac{17}{24 \epsilon ^3}-
\biggl(\frac{149}{48}+\frac{13 \pi ^2}{36}\biggr) \epsilon ^{-2}-
\biggl(\frac{433}{96}+\frac{149 \pi ^2}{72}+\frac{23 \zeta _3}{9}\biggr) \epsilon ^{-1}-
\biggl(-\frac{3817}{64}
\nonumber\\
&+\frac{521 \pi ^2}{144}+\frac{173 \zeta _3}{9}+\frac{341 \pi ^4}{180}\biggr) \epsilon ^0-
\biggl(-\frac{97165}{128}-\frac{9419 \pi ^2}{288}+\frac{1367 \zeta _3}{18}+\frac{1667 \pi ^4}{180}
\nonumber\\
&+\frac{659 \pi ^2 \zeta _3}{27}+\frac{5939 \zeta _5}{15}\biggr) \epsilon ^1-
\biggl(-\frac{4640963}{768}-\frac{80461 \pi ^2}{192}+\frac{1717 \zeta _3}{36}+\frac{833 \pi ^4}{360}+\frac{3527 \pi ^2 \zeta _3}{27}
\nonumber\\
&+\frac{30599 \zeta _5}{15}+\frac{57791 \pi ^6}{5670}+\frac{734 \zeta _3^2}{27}\biggr) \epsilon ^2-
\biggl(-\frac{61900879}{1536}-\frac{410283 \pi ^2}{128}-\frac{55357 \zeta _3}{24}
\nonumber\\
&-\frac{193957 \pi ^4}{720}+\frac{8489 \pi ^2 \zeta _3}{54}+\frac{45833 \zeta _5}{30}+\frac{1106911 \pi ^6}{22680}+\frac{14197 \zeta _3^2}{54}+\frac{21637 \pi ^4 \zeta _3}{135}
\nonumber\\
&+\frac{75407 \pi ^2 \zeta _5}{45}+\frac{522017 \zeta _7}{21}\biggr) \epsilon ^3+
O\left(\epsilon ^4\right)
  \,,
\end{align}
with $s_6=\sum_{m=1}^\infty\sum_{k=1}^{m} (-1)^{m+k}/(m^5k)= 0.98744\ldots.$.

%- }}}
%- {{{ B: alMS al OS

\section*{\label{app::alMS2OS}Appendix B: Relation between $\bar{\alpha}(\mu)$
  and $\alpha$}

In this Appendix we present the result for the relation between the fine
structure constant defined in the $\overline{\rm MS}$ and on-shell
renormalization scheme involving two massive leptons with masses $m_1$ and
$m_2$. We label contributions from leptons with mass $m_1$ and $m_2$ by $n_h$
and $n_l$, respectively. Our result reads
\begin{equation}
\begin{split}
  &\frac{\bar \alpha (\mu)} {\alpha} -1
  =
  \frac{1}{3} l_2 n_l \frac{\alpha}{\pi }
  +\Bigg \{ 
  \frac{1}{9} l_1 l_2 n_h n_l
  +\frac{1}{9} l_2{}^2 n_l{}^2
  +\left(\frac{l_2}{4}+\frac{15}{16}\right) n_l
  \Bigg\}\bigg ( \frac{\alpha}{\pi }\bigg )^2 \\&
  + \Bigg \{
  % nl^3
  \frac{1}{27} l_2{}^3 n_l{}^3 
  % nl^2 nh
  +\frac{1}{9} l_1 l_2{}^2 n_h n_l{}^2
  % nl^2
  + n_l{}^2 \left(\frac{5 l_2{}^2}{24}+\frac{79 l_2}{144}+\frac{7 \zeta_3}{64}+\frac{\pi ^2}{9}-\frac{695}{648}\right) \\
  % nl
  &+ n_l \left(-\frac{1}{3} \pi ^2 a_1-\frac{l_2}{32}+\frac{\zeta_3}{192}+\frac{5 \pi ^2}{24}+\frac{77}{576}\right)
  % nl nh
  \\&
  + n_h n_l \Bigg[ \frac{79 l_1 x^2}{384}-\frac{79 l_2 \left(3
      x^2-8\right)}{1152}+\frac{5 l_1 l_2}{24} 
  +\frac{1}{384} \left(-128 x^4-15 x^2-71\right) \ln ^2(x)
  \\&
  +\frac{1}{3} \left(-x^4+x^3+x-1\right) 
  \text{Li}_2(1-x) 
  +\frac{1}{3} \left(x^4+x^3+x+1\right)
  \text{Li}_2(-x)
  \\&
  +\frac{\left(5 x^6+3 x^4+3 x^2+5\right)}{256 x^3}
  \Big( (
  \text{Li}_2(1-x) 
  + 
  \text{Li}_2(-x) 
  ) \ln (x)
  -2 \text{Li}_3(1-x) 
  - \text{Li}_3(-x) 
  \Big)
  % -\frac{\left(5 x^6+3 x^4+3 x^2+5\right) \text{Li}_3(-x)}{256 x^3}
  % +\frac{\left(5 x^6+3 x^4+3 x^2+5\right) \text{Li}_2(-x)}{256 x^3}
  % 
  \\&
  +\frac{1}{3} \left(x^4+x^3+x+1\right) \ln (x) \ln  (x+1)
  % \\&
  +\frac{\left(5 x^6+3 x^4+3
      x^2+5\right) \ln ^2(x) \ln (x+1)}{512 x^3}
  \\&
  +\frac{405 x^3 \zeta_3+1152 \pi ^2 \left(x^3+x\right)-5994 x^2+243 x \zeta_3-5126}{10368}
  \Bigg ]
  % \\&
  % closing
  \Bigg \} \bigg ( \frac{\alpha }{\pi }\bigg )^3
  \\&
  + \Big \{ n_h \leftrightarrow n_l , m_1 \leftrightarrow m_2 , x \leftrightarrow \frac{1}{x}\Big \} \,,
\end{split}
\end{equation}
with $x=m_1/m_2$, $l_k = \ln (\mu^2 / m_k^2)$ and $a_1 = \ln 2$.

%- }}}

\end{appendix}

%- }}}

%- {{{ bibliography

%- }}}

\end{document}